\documentstyle[11pt,newpasp,twoside,epsf]{article}
\def\edcomment#1{\iffalse\marginpar{\raggedright\sl#1\/}\else\relax\fi}
\reversemarginpar

\begin{document}
\title{A Study of the Coronal Plasma in RS CVn binary systems: HR~1099 and co.}

\author{M. Audard\altaffilmark{1}, %
M. G\"udel\altaffilmark{1}, %
A. Sres\altaffilmark{1,2}, %
R. Mewe\altaffilmark{3}, %
A.~J.~J. Raassen\altaffilmark{3,4}, %
E. Behar\altaffilmark{5}, %
C.~R. Foley\altaffilmark{6}, %
R.~L.~J.~van~der~Meer\altaffilmark{3}
}
\altaffiltext{1}{Paul Scherrer Institut, W\"urenlingen \& Villigen, 5232 Villigen PSI,
Switzerland}
\altaffiltext{2}{Institute of Astronomy, ETH Zentrum, 8092 Z\"urich, Switzerland}
\altaffiltext{3}{Space Research Organization Netherlands, Sorbonnelaan 2, 3584 CA Utrecht, The
Netherlands}
\altaffiltext{4}{Astronomical Institute ``Anton Pannekoek'', Kruislaan 403, 1098
SJ Amsterdam, The Netherlands}
\altaffiltext{5}{Columbia Astrophysics Laboratory, Columbia University, New York, NY 10027,
USA}
\altaffiltext{6}{Mullard Space Science Laboratory, UCL, Holmbury St.
Mary, Dorking, Surrey, RH56NT, UK}

\begin{abstract}
\textit{XMM-Newton} has been performing comprehensive 
studies of X-ray luminous RS CVn binary systems in its calibration and
guaranteed time programs. We present results from ongoing investigations 
in the context of a systematic study of coronal emission from RS CVns. 
We concentrate here on coronal abundances and
investigate the abundance pattern in RS CVn binaries as a function
of activity and average temperature. We find a transition from an Inverse
First Ionization Potential (FIP) effect towards an absence of a clear trend (no FIP)
in intermediately active RS CVn systems. This scheme fits well into the long-term
evolution from an IFIP to a FIP effect found in solar analogs. We further study 
variations in the elemental abundances during a large flare. 
\end{abstract}

\section{Introduction}
RS CVn binary systems generally consist of a chromospherically
active evolved star tidally locked to a main-sequence or subgiant companion
(Hall 1976). Short periods of a few days are typically observed. The empirical
relation between rotation and activity in late-type stars (e.g., Noyes et al. 1984)
then implies high levels of activity, observed as strong emission of 
chromospheric lines and saturated X-ray emission (Dempsey et al. 1993).
The high-resolution X-ray spectra of the brightest and nearby RS CVn binary systems 
obtained by  \textit{XMM-Newton} are well-exposed and
provide a high signal-to-noise ratio. They allow us to study the
rich forest of X-ray lines emitted by elements abundant in the coronae, such as
C, N, O, Ne, Mg, Si, S, Ar, Ca, Fe, and Ni. The profusion of lines observed with the
new high-resolution instruments onboard \textit{XMM-Newton} and \textit{Chandra}
can benchmark atomic databases for completeness and accuracy. Recent results show that models reproduce
fairly well the observed real spectra (e.g., Audard et al. 2001a, Behar, Cottam, 
\& Kahn 2001), although a significant number of lines, mainly from Si, S, Ar, and Ca
L-shell lines, are either absent in the atomic databases or are not properly 
reproduced (Audard et al. 2001a). We have therefore discarded
wavelength regions where such lines dominate in order not to bias the spectral
fitting convergence, and particularly to get more accurate elemental abundances.

Stellar coronal abundances have frequently been determined using the moderate 
spectral resolution of CCD spectra from \textit{ASCA} (e.g., Drake 1996, G\"udel et al. 1999)
or from the low sensitivity spectrometers onboard \textit{EUVE} (e.g., Drake, Laming,
\& Widing 1995, Laming, Drake, \& Widing 1996, Schmitt et al. 1996, Drake, Laming, 
\& Widing 1997). The abundance pattern in stellar coronae is complementary to the 
well-studied, but still puzzling, abundance pattern in the Sun: in brief, the 
solar corona and the solar wind display a so-called ``First Ionization Potential'' (FIP) 
effect, for which the current consensus is that the abundances of low-FIP ($<10$~eV) 
elements are \emph{enhanced} relative to their respective photospheric abundance, 
while the abundances of high-FIP ($>10$~eV) elements are photospheric 
(e.g., Haisch, Saba, \& Meyer 1996). Stellar coronal observations often showed a deficiency of metals relative to the \emph{solar}
photospheric abundances (Schmitt et al. 1996), with (Fe/H)/(Fe/H)$_\odot$
ratios around 10 to 20~\% in
active RS CVn binary systems. \textit{EUVE} spectra either showed
the absence of any FIP-related bias (Drake et al. 1995), or the presence 
(Drake et al. 1997) of a FIP effect 
in inactive stellar coronae. The new X-ray observatories \textit{XMM-Newton} and \textit{Chandra} 
combine the high spectral resolution with moderate effective areas to routinely 
obtain data useful to measure the abundances in stellar 
coronae.

A recent study with the deeply exposed  \textit{XMM-Newton} RGS spectrum of the
RS CVn binary system HR~1099 by Brinkman et al. (2001) showed a trend towards enhanced
high-FIP elemental abundances (normalized to O and relative to the solar
photospheric abundances, Anders \& Grevesse 1989), while low-FIP elemental
abundances are depleted; this effect was dubbed the ``Inverse FIP'' effect.
Subsequent studies of different active stars also showed such a trend (G\"udel et al.
2001ab), while the intermediately active binary Capella showed neither a FIP nor
an IFIP effect (Audard et al. 2001a). However, we emphasize that stellar coronal abundances are  
often normalized to the \emph{solar} photospheric abundances, while they should
better be normalized to the \emph{stellar} photospheric abundances. The latter
are, however, difficult to measure because of the enhanced chromospheric activity, the high 
rotation rate and the presence of spots in active stars, particularly in RS CVn
binaries. Nevertheless, for some stars, photospheric abundances are known. 
G\"udel et al. (2002; also in these proceedings) discuss the transition from an IFIP to 
a normal FIP effect in the long-term evolution of the coronae from active to inactive 
solar analogs; all targets  have photospheric abundances indistinguishable from
those of the Sun, therefore suggesting that the observed transition is real.

The apparent depletion of low-FIP elements during quiescence should be contrasted with
a significant increase of average metal abundances during flares
(e.g., Ottmann \& Schmitt 1996, Favata et al. 2000). With the higher spectral resolution of
\textit{ASCA}, individual elemental abundances were derived during time slices of a 
large flare in UX Ari (G\"udel et al. 1999); low-FIP elements were found to increase
more significantly than the high-FIP elements. Recently, high-resolution X-ray spectroscopy of
a flare in HR~1099 with \textit{XMM-Newton} identified similar behavior
(Audard, G\"udel, \& Mewe 2001b).

In this paper, we present abundance patterns in RS CVn binary systems observed by 
\textit{XMM-Newton}. 
Our results generally show i) a transition from an IFIP effect to an absence of
a FIP bias with decreasing
activity, compatible with a similar transition observed in solar analogs (G\"udel et 
al. 2002; also in these proceedings), ii) a depletion of coronal abundances of low-FIP
elements with increasing average coronal temperature, while high-FIP elemental abundances
stay constant, iii) an enhancement of low-FIP elemental abundances during flares, while
high-FIP elemental abundances stay constant.

\section{Observations and Data Analysis}
\begin{figure}[!t]
\plotone{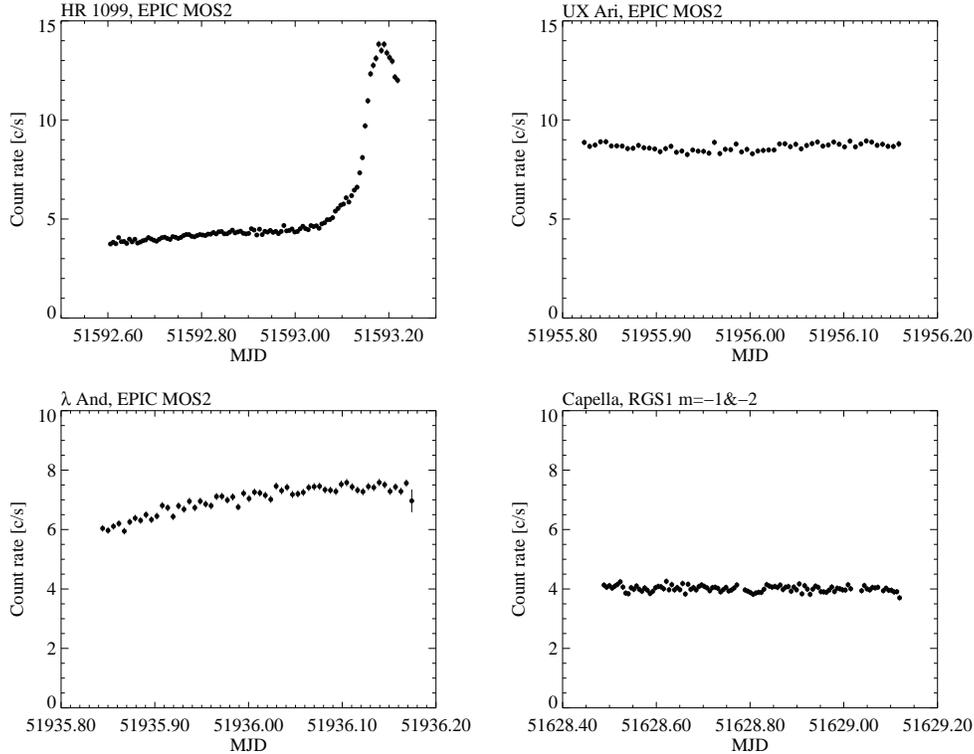}
\caption{X-ray light curves of HR~1099, UX Ari, $\lambda$ And, and Capella. The sensitive
EPIC MOS2 light curves are shown, except for Capella where the sum of the RGS1 first and
second order light curves is given, due to substantial pile-up and
optical contamination of the EPIC data. The bin size is 500~s. Note that for HR~1099, an
annulus extraction region was used to account for pile-up in the
central part of the EPIC PSF. Also, note that only the quiescent
part (MJD $< 51593.074$) of HR~1099 has been included in the analysis; see Fig.~4 for 
the flare analysis (also Audard et al. 2001b).}
\end{figure}

\textit{XMM-Newton} observed several RS CVn binary systems as part of the RGS
stellar Guaranteed
Time Program. In this paper, we will concentrate on the quiescent observations of
HR~1099, UX Ari, $\lambda$~And, and Capella. Their light curves are shown in Figure~1.
We have fitted simultaneously the RGS1, RGS2, and EPIC MOS2 spectra (except for Capella
where no EPIC data are available) in XSPEC 11.0.1aj (Arnaud 1996) using the
\texttt{vapec} model (APEC code with variable abundances). We have removed significant
parts of the RGS spectra above 20~\AA\  to take into account the inaccuracy and
incompleteness of the atomic database for non-Fe L-shell atomic transitions.
Additionally, some Fe L-shell lines with inaccurate atomic data were not fitted. A free multiplicative 
constant model has been introduced for cross-calibration uncertainties, finite
extraction region and, in the case of HR~1099, the annulus-shaped extraction region.
Notice that the RGS1 and RGS2 each suffer from the loss of one CCD, but the
combined spectra cover the whole RGS wavelength range (see den Herder et al.
2001). We have used either i) a
multi-temperature (5T) approach with free T, emission measures (EM) and abundances 
(the latter linked between the components), or ii) 10 components on a grid of
fixed T, but free EM 
and free abundances (linked between the components). Both methods proved to give
reasonable fits with similar abundances. Therefore we will
only give results from the 10-T approach.

\section{Results}

\begin{figure}[!t]
\plotone{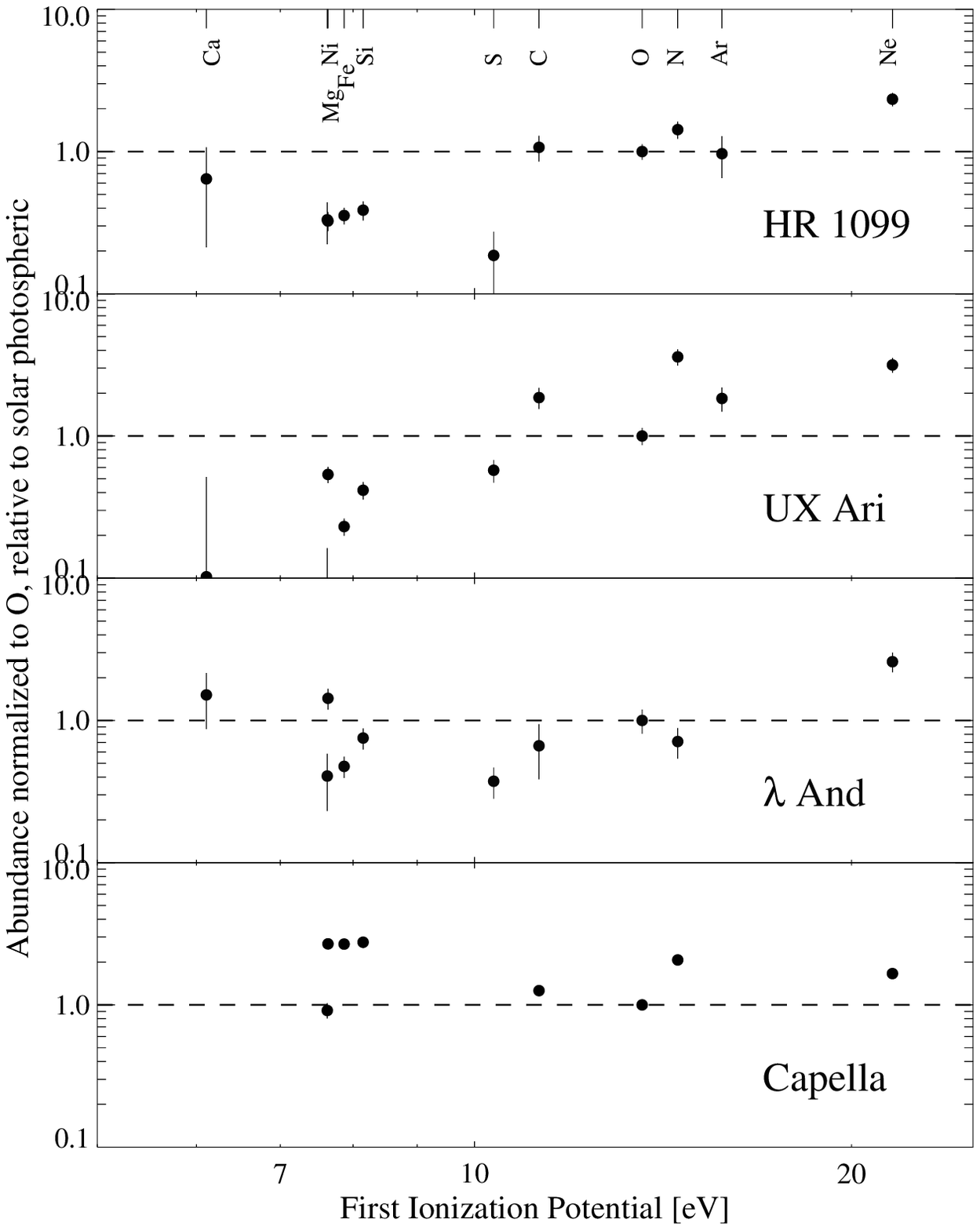}
\caption{Coronal abundance ratios in RS CVn binaries as a function of FIP. The abundances have been
normalized to O to allow for comparison, hence $\mathrm{M/O = \{(M/H)/(M/H)_\odot\} / \{
(O/H)/(O/H)_\odot \}}$. Solar photospheric abundances from Anders \& Grevesse (1989) 
have been used, except for Fe (Grevesse \& Sauval 1999).}
\end{figure}

Figure~2 shows the coronal abundances normalized to the O abundance in order to ease
comparison between the different stars. Note that the solar photospheric abundances from Anders \&
Grevesse (1989) were used, except for the Fe abundance that was taken from Grevesse \&
Sauval (1999). The panels are ordered in decreasing activity from top to
bottom. A clear trend for high-FIP M/O ratios to significantly exceed
low-FIP M/O ratios is seen in the very active stars (HR~1099, UX Ari), while for the less
active, not tidally locked $\lambda$ And, no clear correlation is observed.
Finally, in the
intermediately active Capella, any FIP bias is absent with a possibility for a
weak FIP effect.

The ``average'' temperature of a stellar corona is an indicator of activity. We
define here $\log <T> = (\Sigma_i \log T_i \times EM_i) /(\Sigma_i EM_i)  $,
where $T_i$ and $EM_i$ are
the 10-T model temperatures and emission measures, respectively. The abundance ratios
relative to O have been plotted against $<T>$. Figure~3 shows examples for the low-FIP
element Fe and for the high-FIP element Ne. Data points from solar analogs (G\"udel et
al. 2002 and in these proceedings) have been added; both panels show a very different behavior
in the abundance ratios: while the Fe/O ratios exponentially decrease with increasing
temperature, the Ne/O ratios show no correlation with the average coronal temperature.
Other low-FIP elements (e.g., Mg, Si) show a similar trend.

\begin{figure}[!t]
\plottwo{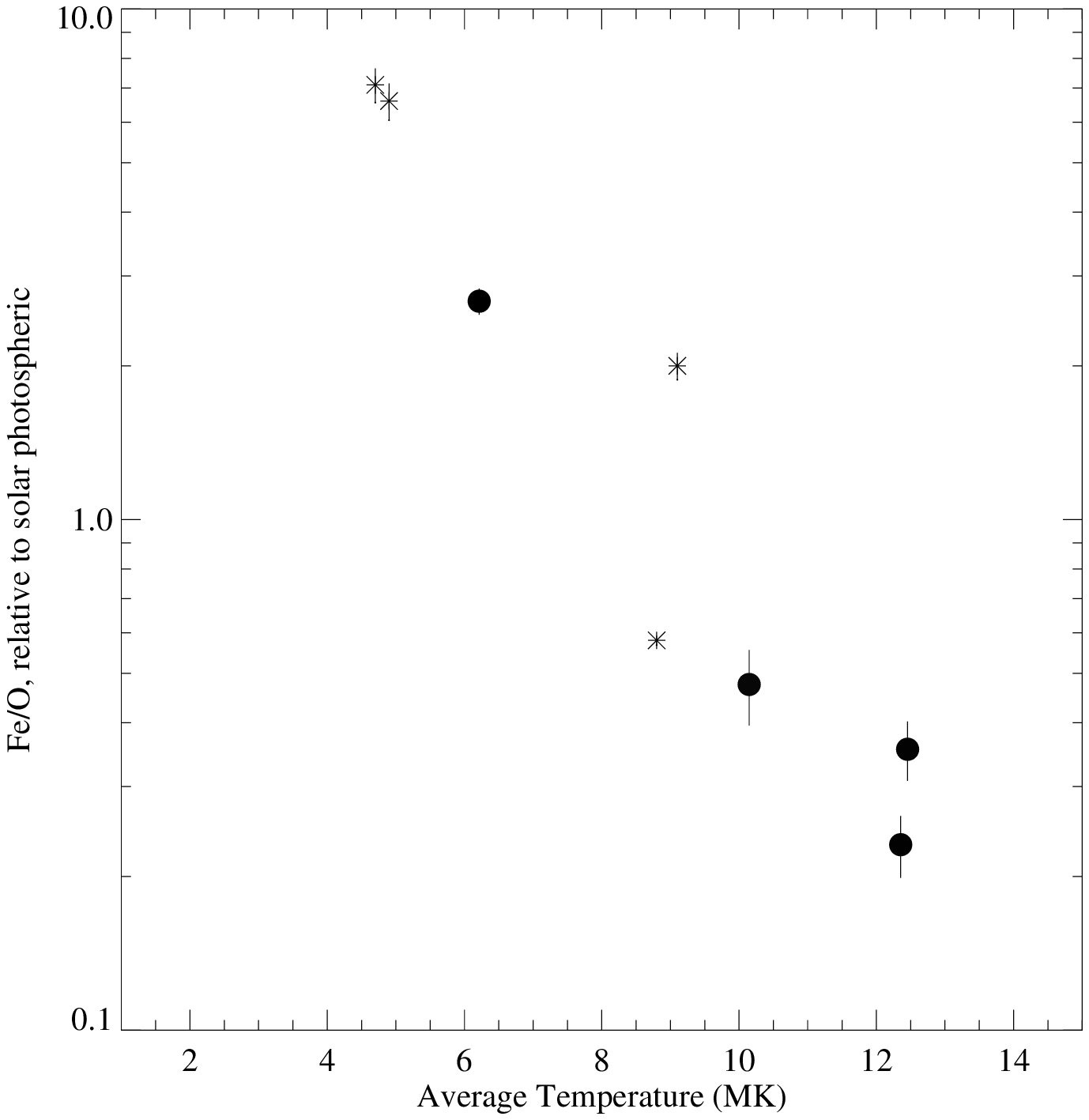}{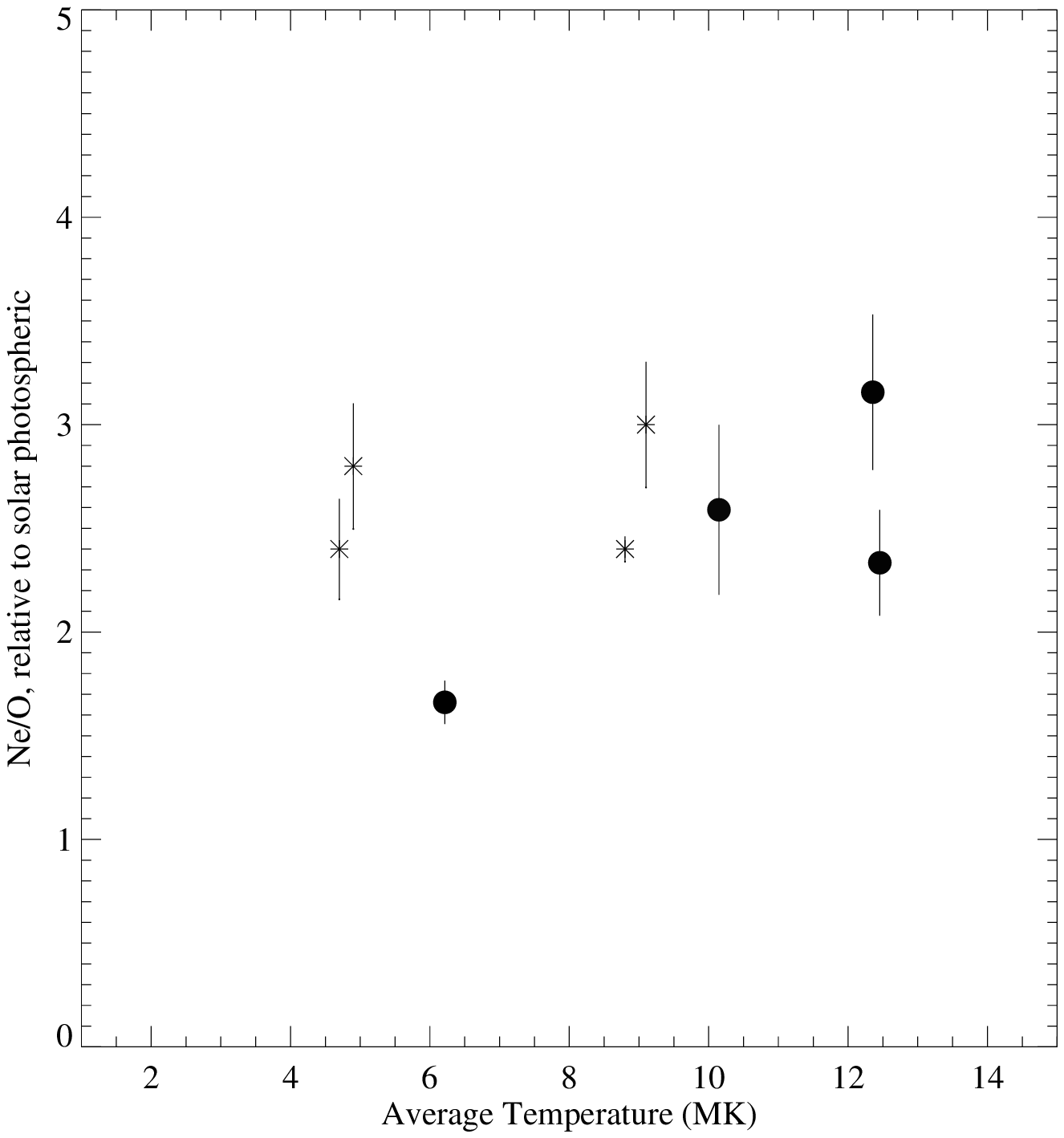}
\caption{Coronal abundance ratio as a function of average coronal temperature. 
\textit{Left}: The Fe/O ratios (relative to solar photospheric) for RS CVn binary systems
(dots) together with similar ratios for solar analogs (stars, from G\"udel et al. 2002; 
also these proceedings). Note the logarithmic vertical scale. \textit{Right}: Similar but for the Ne/O ratios. 
Notice that other low-FIP
elements show a similar trend as Fe and other high-FIP elements show a
similar but less clear constant behavior.}
\end{figure}

The above trends are given for ``quiescent'' X-ray coronae; data from previous satellites
showed that the metallicity Z, or the Fe abundance, generally increases during large
flares. Higher spectral resolution data showed that low-FIP elements 
increased more significantly than the high-FIP elements (G\"udel et al. 1999). 
We reanalyzed the HR~1099 flare (previously published by Audard et al. 2001b) 
applying a more recent calibration. Figure~4 shows the Fe/O and Ne/O ratios as 
a function of the average temperature during quiescence, flare rise, and flare 
peak.

\begin{figure}[!t]
\plottwo{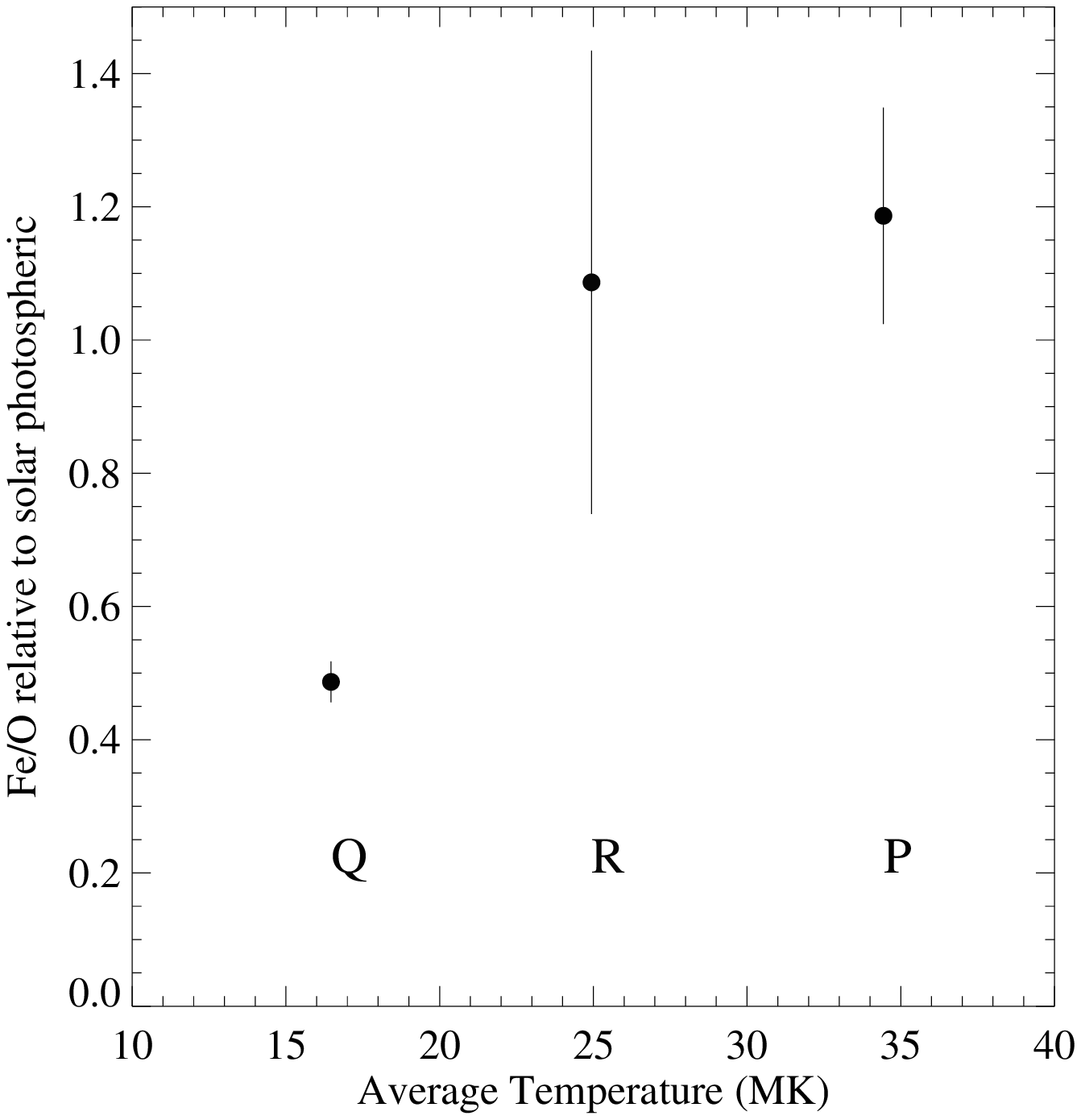}{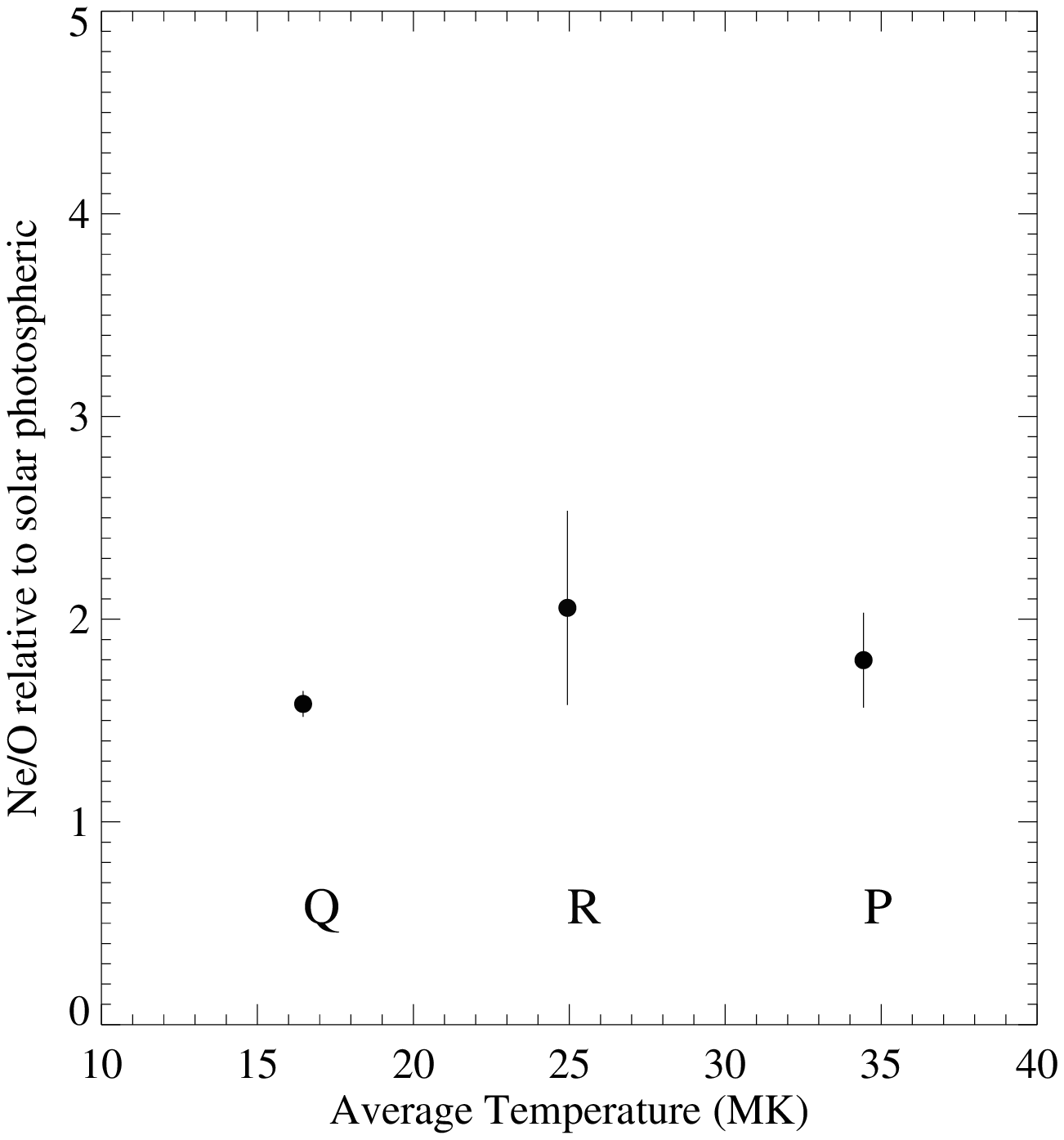}
\caption{Coronal abundance ratio as a function of average coronal temperature during
an HR~1099 flare. Left panel give the Fe/O ratios, while the right panel gives the Ne/O
ratios. `Q' stands for quiescent, `R' for flare rise, and `P' for flare peak.}
\end{figure}

Consistently with our previous analysis (Audard et al. 2001b) and with previous 
results, the Fe/H abundance increases during the flare; however the Ne/H 
abundance stays constant. After normalization with the O abundance, the Fe/O
ratio still increases and Ne/O still remains
constant, consistent with the picture that low-FIP elements are enhanced during flares
while high-FIP elements stay equally abundant. This behavior of
other low-FIP elements and high-FIP elements is analogous.

\section{Discussion and Conclusion}

Our \textit{XMM-Newton} results show a correlation between the abundance
pattern in active stars and the activity level. The bright and very active RS
CVn binary systems show a coronal abundance enhancement for 
high-FIP elements (e.g., C, N, O, Ne) compared to low-FIP elements
(e.g., Fe, Mg, Si). We interpret this result as further evidence for an
``Inverse First Ionization Potential'' effect such as found by Brinkman et al.
(2001) in a deep exposure of HR~1099. Our sample, containing HR~1099, UX
Ari, $\lambda$ And, and Capella, covers high to intermediate activity levels. 
The slope of the abundance ratios (M/O) as a function of the FIP curve
decreases with decreasing activity, with our ``least active'' RS CVn
system (Capella) suggesting either an absence of a FIP bias or a weak FIP effect
(Fig.~2). Similarly, using the average coronal temperature as an activity
indicator, we show that low-FIP elemental abundances decrease with increasing 
temperature, while the high-FIP abundances stay
constant. Note that evidently this behavior would change if abundances were 
normalized to a low-FIP element such Fe. However, absolute abundances suggest
that \emph{only} the low-FIP elements are sensistive to the activity level,
while this is not the case for the high-FIP elements.

Such correlations correspond well to the long-term evolution from an IFIP
effect to a normal FIP effect in solar analogs (G\"udel et al. 2002; also
these proceedings). These latter stars are of solar photospheric composition. 
For  most RS CVn binary systems, photospheric abundances are unknown,
and when known, there is a large scatter in their measurements because they
are difficult to derive from optical spectroscopy mainly
due to chromospheric activity, high rotation velocities, and the presence of
spots on the stellar surface. $\lambda$ And is one of the rare cases with
several measured photospheric abundances. We used the \emph{stellar}
abundances derived by Donati, Henry, \& Hall (1995): a similar distribution as
in Fig.~2 is obtained. Notice that it is pivotal to measure accurate
photospheric abundances in the most active stars that show a clear IFIP effect
(e.g., HR 1099, UX Ari) and compare them with coronal abundances. It will allow
us to verify whether low-FIP elements really are depleted in the corona of the most
active stars.

\acknowledgments
M.~A. acknowledges support from the Swiss National 
Science Foundation (grant 2100-049343). He also thanks the
organizers of the Stellar Coronae conference for financial
support. The Space Research Organization of the Netherlands (SRON) is supported 
financially by NWO. This work is based on observations obtained with XMM-Newton, an ESA science 
 mission with instruments and contributions directly funded by ESA Member 
 States and the USA (NASA).

\end{document}